\documentclass[aip,reprint]{revtex4-1}
\usepackage{amsmath}
\usepackage{amssymb}
\usepackage{hyperref}
\hypersetup{colorlinks=true,linkcolor=blue,citecolor=blue,urlcolor=blue}
\usepackage{graphicx}
\usepackage{siunitx}
\DeclareSIUnit{\angstrom}{\text{\AA}}
\usepackage[normalem]{ulem}
\usepackage{braket}
\usepackage[utf8]{inputenc}
\usepackage[T1]{fontenc}
\usepackage{bm}
\begin{document}

\title{Constraints on Atomistic Disorder for Scalable Electron Spin Shuttling}
\thanks{Copyright 2026 Prentki, Philippopoulos, Mostaan, and Beaudoin. This article is distributed under a Creative Commons Attribution-NonCommercial-NoDerivs 4.0 International (CC BY-NC-ND) License (\url{https://creativecommons.org/licenses/by-nc-nd/4.0/}). This article appeared in \emph{Applied Physics Letters} \textbf{129}, 054001 (2026) and may be found at \url{https://doi.org/10.1063/5.0345368}.}
\author{Rapha\"{e}l J. \surname{Prentki}}
\email{rprentki@nanoacademic.com}
\author{Pericles \surname{Philippopoulos}}
\author{Mohammad Reza \surname{Mostaan}}
\author{F\'{e}lix \surname{Beaudoin}}
\affiliation{Nanoacademic Technologies Inc., Montr\'{e}al, Qu\'{e}bec H3A 1E7, Canada}
\date{\today}

\begin{abstract}
Electron spin shuttling---the gate-controlled, coherent transport of electrons between qubit registers---increases qubit connectivity and enables efficient quantum-error-correction schemes. It is emerging as a key enabler of scalable silicon spin-qubit quantum computing. As an electron travels over micrometers, it encounters angstrom-scale disorder, causing fluctuations in its confinement potential, valley splitting, and valley phase. These lead to leakage into the valley-excited state, limiting high-fidelity shuttling speeds. Accurate predictions of shuttling fidelities thus require modeling tools that link atomistic and mesoscopic physics. We develop a multiscale simulation workflow to quantify these effects in the experimentally-realized Si/SiGe ``QuBus'' conveyor-belt architecture. First, we resolve the time-dependent, gate-controlled device electrostatics by solving the Poisson equation using the finite-element method. Second, we construct conveyor belt atomic structures with realistic atomistic disorder (random alloying and interface roughness); we resolve strain atomistically using the Keating valence force-field model. Third, we perform position-tracked atomistic tight-binding simulations of the shuttled electrons to obtain their time-dependent valley splittings and phases. Finally, these time traces parametrize a time-dependent Schr\"{o}dinger equation, which we solve to predict valley dynamics. We find that interface roughness strongly suppresses shuttling fidelities, with a sharp anomaly near the atomic-layer scale. Overall, our predictions set practical, quantitative guidelines to realize scalable, high-fidelity shuttling in silicon spin-qubit architectures.
\end{abstract}

\maketitle

Spin qubits in gate-defined silicon quantum dots (QD) are a leading platform for dense, scalable quantum computing owing to (1) their compatibility with advanced semiconductor manufacturing processes, which enables device pitches at the \SI{100}{\nano\meter} scale~\cite{zwerver2022qubits}, (2) their high operating temperatures~\cite{yang2020operation}, and (3) their long coherence times and fast, high-fidelity gate operations~\cite{zwanenburg2013silicon,stano2022review}. Standard two-qubit gates based on the exchange coupling require wavefunction overlap and thereby restrict qubit connectivity to nearest neighbors. However, novel schemes for low-overhead, fault-tolerant quantum computing require long-range qubit connectivity~\cite{xu2024constant,bravyi2024high}. While millimeter-scale cavities can mediate distant coupling~\cite{borjans2020resonant}, their large footprints hinder dense integration. Electron spin \emph{shuttling}---the coherent, gate-controlled transport of the charge, orbital, and spin degrees of freedom of an electron over micrometers, between qubit registers---is emerging as a promising alternative~\cite{langrock2023blueprint}.

Two main modes of shuttling are being explored: bucket brigade and conveyor belt (CB). In bucket-brigade shuttling, an electron is sequentially tunneled along a chain of QDs. In CB shuttling, an electron is transported by a smoothly propagating potential well. While bucket-brigade shuttling over \SI{1}{\micro\meter}~\cite{mills2019shuttling} and high-fidelity back-and-forth transfer of an electron between two QDs~\cite{yoneda2021coherent} have been demonstrated, the bucket brigade mode is less scalable due to its length-dependent calibration overhead and signal fan-out problems~\cite{langrock2023blueprint}. By contrast, high-fidelity CB shuttling has been demonstrated using only four control signals~\cite{seidler2022conveyor,xue2024si} and over \SI{10}{\micro\meter}, with an average shuttling speed of \SI{50}{\meter\per\second}~\cite{de2025high}. Furthermore, in the same Si--SiGe device, CB shuttling has been shown to have a spin coherence that is an order of magnitude better than bucket-brigade shuttling~\cite{de2025high}.

CB shuttling is subject to several error channels, notably charge-transfer failure---which occurs when the propagating potential well is distorted by charge defects~\cite{xue2024si,jeon2025robustness,nagai2025digital}---and spin dephasing---which arises in the presence of fluctuating electric and magnetic fields~\cite{de2025high,jeon2025robustness}. Crucially, it is leakage into the \emph{valley}-excited state that sets the practical limit for shuttling speed~\cite{losert2024strategies,david2024long,volmer2025reduction}. While bulk silicon has six degenerate conduction-band valleys, strain and quantum confinement lift this degeneracy in silicon QDs, with the two $\pm z$ valleys dominating device physics~\cite{friesen2007valley}; here, $z$ is the crystal growth axis, assumed to be parallel to the $[001]$ crystal direction. The ground and first-excited (or valley-excited) states of such a QD are thus superpositions of the $\pm z$ valley states; the energy difference between these two eigenstates is known as the valley splitting, and the relative phase between the coefficients of the two valley states in the superpositions is known as the valley phase. As an electron is shuttled, it samples varying atomistic disorder, leading to fluctuating valley splitting and valley phase~\cite{paquelet2022atomic,losert2023practical,lima2024valley,volmer2024mapping,klos2024atomistic}. The time derivative of the valley phase drives leakage into the valley-excited state~\cite{friesen2007valley,saraiva2009physical}; therefore, for a fixed atomistic disorder landscape, the valley leakage increases with shuttling speed. We note that since both shuttling and dense qubit arrays are sensitive to atomic-scale disorder, studying its impact on shuttling provides insight into disorder-related limitations in large-scale qubit architectures.

The main heterostructures under investigation for industrial silicon qubits are Si--SiGe and Si--SiO$_2$~\cite{zwanenburg2013silicon,stano2022review}. A major contributor to atomistic disorder in the former system is random alloying, namely the random distribution of Ge atoms and random local stoichiometry of Si$_{1-x}$Ge$_x$ layers. Another major contributor in both systems is interface roughness: the interface between the Si well and its SiGe or SiO$_2$ barrier(s) is typically not pristine but rather exhibits height fluctuations that depend on various factors, such as the heterostructure growth techniques and conditions. The impact of random alloying on shuttling fidelities has been investigated through statistical (continuum) models of valley splitting fluctuations, enabling shuttling fidelity optimizations via vertical electric fields and/or trajectory shaping~\cite{david2024long,losert2024strategies,volmer2025reduction}. Furthermore, the root mean square (RMS) of height fluctuations of rough Si--SiGe and Si--SiO$_{2}$ interfaces falls between \SI{1} and \SI{10}{\angstrom}~\cite{pena2024modeling,paquelet2022atomic,cifuentes2024bounds}, which is known to lead to significant fluctuations in valley splitting and valley phase~\cite{paquelet2022atomic,cifuentes2024bounds}. Crucially, the impact of interface roughness on shuttling fidelities has yet to be quantified through explicit atomistic disorder in realistic device geometries.

In this Letter, we aim to quantify constraints on atomistic disorder for practical, scalable shuttling. Reliable predictions of shuttling fidelities require multiscale simulations that link angstrom-scale disorder to micrometer-scale shuttling distances. To do so, we develop a state-of-the-art simulation workflow that resolves device electrostatics at the continuum level and disorder, strain, and valley physics at the atomistic level, as implemented in the QTCAD\textsuperscript{\tiny\textregistered} package~\cite{qtcad}. Crucially, we find that shuttling fidelities exhibit a non-monotonic dependence on interface roughness RMS height: they collapse when the RMS height is comparable to a single atomic layer ($\approx \SI{1}{\angstrom}$), while, excluding this outlier, they sharply decrease with increasing RMS height.

\begin{figure}[!tbp]
\includegraphics[width=\columnwidth]{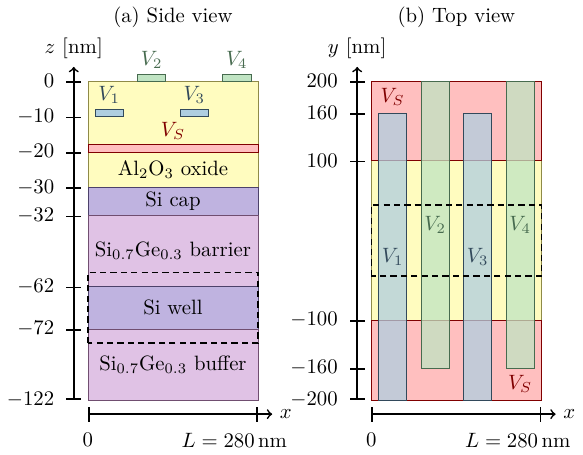}
\caption{Schematic of the simulated CB device, which models the ``QuBus'' of Ref.~\onlinecite{seidler2022conveyor} (not to scale). The clavier gates ($V_{1,2,3,4}$) are \SI{60}{\nano\meter} wide and separated by \SI{10}{\nano\meter} gaps. The finite-element simulations cover the structure shown in its entirety, while the atomistic simulations are restricted to the region inside the dashed lines.}
\label{fig:diagram}
\end{figure}

As a vehicle for this study, we consider the experimentally realized CB device of Ref.~\onlinecite{seidler2022conveyor}, known as ``QuBus,'' on which an electron charge shuttling fidelity of $99.7\pm0.3\%$ was observed over \SI{19}{\micro\meter}~\cite{xue2024si}. Electrons are shuttled through a \SI{10}{\nano\meter}-thick Si well interposed between Si$_{0.7}$Ge$_{0.3}$ layers, which impart biaxial tensile strain on the well. This SiGe--Si--SiGe heterostructure is topped by an Si cap and an Al$_2$O$_3$ oxide layer, in which six gates are embedded [Fig.~\ref{fig:diagram}]. A constant voltage $V_S$ is applied to the two screening gates to confine electrons along the $y$ axis. Sinusoidal voltages are applied on the four clavier gates to shuttle electrons from left to right,
\begin{align}
\nonumber
V_{j}(t) &= A_S \cos\left[\frac{2\pi t}{T} - \frac{\pi}{2} (j-1)\right] \\
&+ B_S + \Delta B_S \left[(j+1) \bmod 2\right]\,,
\label{eq:V1234}
\end{align}
where $j\in\{1,2,3,4\}$, $t$ is time, $T$ is the voltage period, $A_S$ is the voltage amplitude, $B_S$ is a common voltage offset, and $\Delta B_S$ is an additional offset for gates $2$ and $4$ to account for the fact that they are further from the Si well. The simulations presented in this Letter assume $A_S=\SI{60}{\milli\volt}$, $\Delta B_S=\SI{10}{\milli\volt}$, $V_S - B_S = - \SI{150}{\milli\volt}$, and $B_S - \frac{W}{e} = - \SI{4.01}{\volt}$ ($W$ and $e$ are, respectively, the gate metal work function and the elementary charge), chosen to be representative of the voltage ranges explored experimentally in Ref.~\onlinecite{seidler2022conveyor}.

While the experimental device includes single-electron transistors to inject and detect electrons in the CB, our simulation is restricted to a single unit cell of the CB. As such, we impose periodicity along the shuttling axis ($x=0$ and $x=L$ are equivalent) to model an infinitely long CB. Overall, our simulation model replicates the experimental CB in terms of material composition, geometry, and applied voltages.

The first step of the multiscale simulation workflow is to obtain the electric potential $\varphi(\mathbf{r},t)$ as a function of position $\mathbf{r}$ and time $t$. Given the absence of doping and classical charge carriers under shuttling conditions, $\varphi(\mathbf{r},t)$ satisfies the linear Poisson equation~\cite{beaudoin2022robust,prentki2023robust,philippopoulos2024analysis},
\begin{equation}
\bm{\nabla}\cdot\left[\varepsilon(\mathbf{r})\bm{\nabla}\varphi(\mathbf{r},t)\right]=0\,,
\label{eq:poisson}
\end{equation}
where $\varepsilon(\mathbf{r})$ is the permittivity, subject to periodic boundary conditions (BCs) along $x$, Dirichlet BCs set by the applied gate voltages on the gate surfaces, and homogeneous Neumann BCs on all other surfaces of the simulation domain. Exploiting linearity, we obtain $\varphi(\mathbf{r},t)$ for arbitrary $t$ from six static solutions of Eq.~\ref{eq:poisson}, one per gate, following Ref.~\onlinecite{jeon2025robustness}. The finite-element mesh contains $3.4\times 10^6$ nodes and covers the structure illustrated in Fig.~\ref{fig:diagram} in its entirety.

\begin{figure}[!tbp]
\includegraphics[width=\columnwidth]{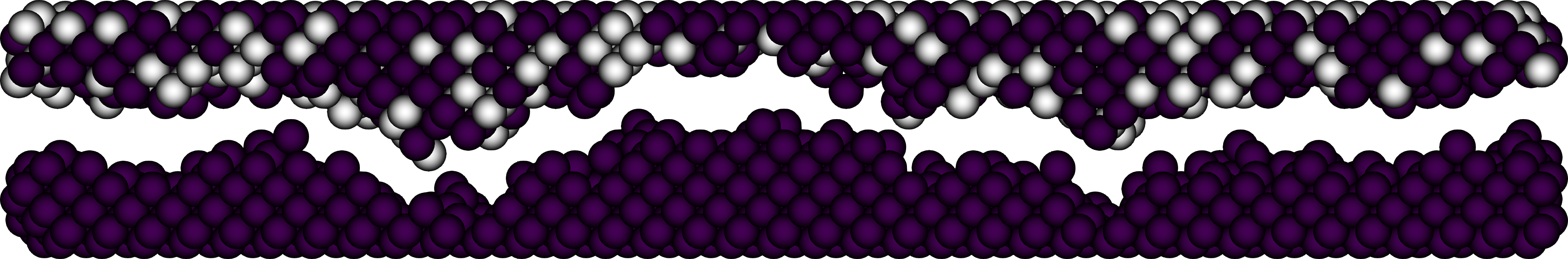}
\caption{Rough Si--SiGe interface with RMS height of \SI{3}{\angstrom}. Purple (white) spheres denote Si (Ge) atoms. The Si and SiGe layers are vertically offset for clarity.}
\label{fig:rough}
\end{figure}

Next, we construct an atomic structure for the CB. The Si--SiGe interfaces may be characterized by their 2D surface roughness power spectrum, $C_{\mathrm{2D}}(\mathbf{k})=1/(2\pi)^2\iint\left<h(\mathbf{s})h(\mathbf{0})\right>e^{i\mathbf{k}\cdot\mathbf{s}}d\mathbf{s}$, where $h(\mathbf{s})$ denotes the interface height at $\mathbf{s}=(x,y)$. Experimentally, interfaces are typically isotropic and self-similar, yielding~\cite{rodriguez2025determine}
\begin{equation}
C_{\mathrm{2D}}(\mathbf{k})\propto |\mathbf{k}|^{-2(1+H)}\,,
\label{eq:C2D}
\end{equation}
where $H$ is the Hurst exponent; here, we assume a typical value of $H=0.3$~\cite{cifuentes2024bounds}. After interface characterization, e.g., via x-ray/neutron scattering~\cite{sinha1988x} or transmission electron microscopy~\cite{cifuentes2024bounds}, Eq.~\ref{eq:C2D} may be inverted to generate a random rough interface with the desired statistics~\cite{iteney2024pyrough,rodriguez2025determine},
\begin{equation}
h(\mathbf{s})=\left<h\right>+C\sum_{n=1}^N G_n |\mathbf{k}_n |^{-(1+H)}\cos(\mathbf{k}_n\cdot\mathbf{s}+U_n)\,,
\label{eq:rough}
\end{equation}
where $C$ is a normalization constant that sets the RMS height and the sum runs over $N$ randomly sampled wavevectors $\mathbf{k}_n$, each of which is associated with a random weight $G_n$, sampled from a standard normal distribution, and a random phase $U_n$, sampled uniformly on $[0,\pi]$. Self-similarity holds only over a finite lengthscale range $[\lambda_{\mathrm{min}},\lambda_{\mathrm{max}}]$; here, we assume typical values of $\lambda_{\mathrm{min}}=\SI{1}{\angstrom}$ and $\lambda_{\mathrm{max}}=\SI{100}{\nano\meter}$~\cite{pena2024modeling,cifuentes2024bounds}. As such, the wavevectors $\mathbf{k}_n$ are randomly sampled on an annulus of inner radius $2\pi/\lambda_{\mathrm{max}}$ and outer radius $2\pi/\lambda_{\mathrm{min}}$. An example of a rough Si--SiGe interface generated via this procedure is shown in Fig.~\ref{fig:rough}.

The atomic structure contains disorder in the form of random alloying and rough interfaces, which leads to a nontrivial strain field. Strain is known to strongly impact the electronic structure of quantum dots~\cite{boykin2004valley}; resolving the strain field atomistically is therefore essential. To do so, we employ the Keating valence force-field model, wherein an elastic energy functional is minimized with respect to the atomic positions to obtain a relaxed atomic structure~\cite{keating1966effect,niquet2009onsite}. The atomic structure spans the full length of the CB ($0<x<L$) and includes the Si well and \SI{12}{\nano\meter} of the adjacent SiGe layers; it contains $1.3\times 10^7$ atoms. We note that atom probe tomography measurements indicate that Si--SiGe interfaces may exhibit finite Ge concentration gradients~\cite{paquelet2022atomic,klos2024atomistic}. We assume an atomically abrupt profile to isolate the impact of interface roughness. While a smooth gradient may modify confinement and valley splitting magnitudes, it is not expected to qualitatively alter the disorder-driven mechanisms discussed here.

\begin{figure*}[!tbp]
\includegraphics[width=\textwidth]{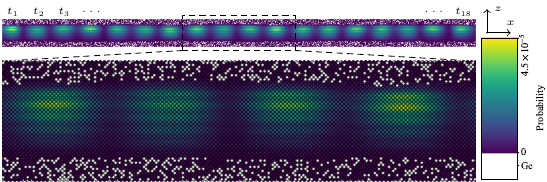}
\caption{Snapshots of the shuttled electron ground state probability density, $|\psi_0(\mathbf{r},t_{p})|^2$ for $p\in\{1,2,\ldots,18\}$, overlaid onto a slice of the simulated CB atomic structure, for a system with rough Si--SiGe interface RMS height of \SI{2}{\angstrom}. The slice contains the three atomic layers closest to the $y=0$ plane and spans the full length of the simulated CB, from $x=0$ to $x=L$. Each sphere represents an atom. Ge atoms are shown in white. The color of each Si atom encodes the probability of finding the electron on that atom. Movies available online~\cite{arxivmovies}.}
\label{fig:TB}
\end{figure*}

\begin{figure}[!tbp]
\includegraphics[width=\columnwidth]{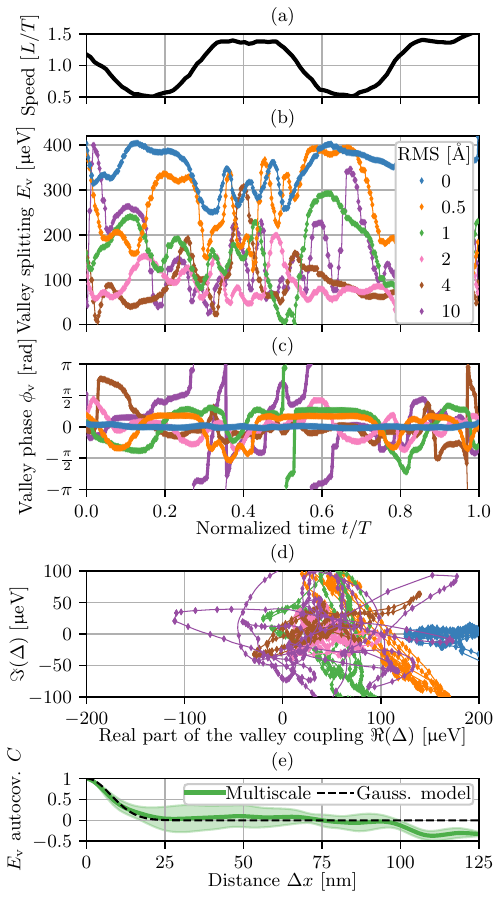}
\caption{(a) Time-dependent electron speed along the shuttling axis $x$ obtained from finite-element electrostatics [Eq.~\ref{eq:poisson}]. (b)--(c) Time-dependent (b) valley splitting and (c) valley phase of CBs with rough Si--SiGe interface RMS heights ranging from $0$ to $\SI{10}{\angstrom}$. Time is normalized by the period $T$ of the clavier gate voltages [Eq.~\ref{eq:V1234}]. Diamonds represent the results of TB simulations, while curves are piecewise cubic Hermite interpolating polynomial fits to the data. (d) Complex-plane time traces of the valley coupling integral $\Delta = \frac{E_{\mathrm{v}}}{2}e^{-i\phi_{\mathrm{v}}}$. (e) Valley splitting auto-covariance function $C(\Delta x)\propto\left<\left(E_{\mathrm{v}}(x)-\left<E_{\mathrm{v}}\right>\right)\left(E_{\mathrm{v}}(x+\Delta x)-\left<E_{\mathrm{v}}\right>\right)\right>$ in CBs with rough Si--SiGe interface RMS height of \SI{1}{\angstrom}. The shaded area indicates the variability of the auto-covariance function, corresponding to $\pm 1$ standard deviation computed from an ensemble of six CB atomic structures. The dashed line corresponds to the model $C(\Delta x)\propto e^{-\left(\frac{\Delta x}{\ell}\right)^2}$ (here, $\ell=\SI{11}{\nano\meter}$) expected from the effective-mass approximation~\cite{david2024long}.}
\label{fig:speedvsvptracescov}
\end{figure}

We now extract the electronic structure of the shuttled electron as a function of time. This requires tracking the position of the electron as it is shuttled, which we estimate, for time $t$, from the maximum over position $\mathbf{r}$ of the electric potential $\varphi(\mathbf{r},t)$ [Eq.~\ref{eq:poisson}]; the time derivative of this position gives the electron speed [Fig.~\ref{fig:speedvsvptracescov}(a)]. We express the system's Hamiltonian within the atomistic Slater--Koster tight-binding (TB) model~\cite{slater1954simplified}; we use the $sp^3d^5s^{\star}$ parametrization of Ref.~\onlinecite{niquet2009onsite}, which accurately models the bandstructures of arbitrarily-strained Si, Ge, and SiGe systems. To capture the confinement potential, we interpolate $\varphi(\mathbf{r},t)$ obtained via the finite-element solution of Eq.~\ref{eq:poisson} onto the atomic positions and modify the on-site energies of the TB Hamiltonian accordingly. The electronic wavefunctions are strongly localized; we thus restrict the TB simulation to a box of $\SI{20}{\nano\meter}\times\SI{20}{\nano\meter}\times\SI{12}{\nano\meter}$, amounting to $2.4\times 10^5$ atoms. The instantaneous electronic structure is then resolved via an efficient sparse eigensolver. We repeat this procedure over $420$ time steps, which track the electron from $x=0$ to $x=L$; this amounts to one time step every $4.9$ atomic layers, on average. Figure~\ref{fig:TB} shows the electron ground state at a few time steps obtained via this procedure (movies available online~\cite{arxivmovies}); wavefunction deformations due to atomistic disorder are visually evident. We note that these $420$ TB calculations constitute the computational bottleneck of our workflow, with a cumulative computation time of $3.4$ days on a $64$-core processor.

From these position-tracked TB simulations, we obtain the instantaneous eigenenergies $E_{0}(t)$ and $E_{1}(t)$ of the ground and valley-excited states, with corresponding eigenfunctions $\psi_{0}(\mathbf{r},t)$ and $\psi_{1}(\mathbf{r},t)$. The valley splitting is defined as $E_{\mathrm{v}}(t) = E_{1}(t) - E_{0}(t)$. Within the effective mass approximation~\cite{friesen2007valley},
\begin{equation}
\psi_{m}(\mathbf{r},t) \approx \frac{F(\mathbf{r},t)}{\sqrt{2}} \sum_{\nu=\pm 1} \nu^{1-m} u_{\nu z}(\mathbf{r})e^{i\nu\left[k_0 z + \frac{1}{2}\phi_{\mathrm{v}}(t)\right]}\,,
\end{equation}
where $m\in\{0,1\}$, $F(\mathbf{r},t)$ is the envelope function, $u_{\pm z}(\mathbf{r})$ are the Bloch amplitudes associated with the $\pm z$ valleys, $k_0=0.84 \frac{2\pi}{a_0}$ is the wavenumber associated with the $\pm z$ valleys, $a_0=\SI{5.43}{\angstrom}$ is the lattice constant of bulk Si, and $\phi_{\mathrm{v}}(t)$ is the valley phase. Considering the density difference $|\psi_0(\mathbf{r},t)|^2-|\psi_1(\mathbf{r},t)|^2$, averaging over the $xy$ plane suppresses lattice-periodic contributions from the Bloch amplitudes, yielding a signal proportional to $\cos[2k_0 z-\phi_{\mathrm{v}}(t)]$. We then perform a Fourier transform along $z$, the argument of which defines $\phi_{\mathrm v}(t)$ when evaluated at wavenumber $2k_0$~\cite{cifuentes2024bounds}. The resulting valley splitting and phase are shown in Fig.~\ref{fig:speedvsvptracescov}(b)--(c). As expected from the $2k_0$ theory~\cite{losert2023practical}, the mean of the valley splitting decreases with increasing Si--SiGe interface roughness; both the valley splitting and valley phase exhibit greater relative fluctuations with increased roughness. In addition, when the RMS height is close to the atomic layer thickness $a_0/4\approx\SI{1}{\angstrom}$, single-monolayer steps in the Si--SiGe interfaces are likely. When the shuttled electron crosses such a step, the valley coupling integral~\cite{friesen2007valley,losert2023practical} acquires nearly opposite phase contributions from adjacent terraces, yielding destructive interference and near-zero valley splitting; we note that Ge atom diffusion across Si--SiGe interfaces would reduce the prevalence of single-monolayer steps. In addition, as shown in Fig.~\ref{fig:speedvsvptracescov}(d), the complex-plane time traces of the valley coupling integral exhibit a nonzero offset for small RMS heights, reflecting the deterministic valley splitting of heterostructures with sharp interfaces. By contrast, as the interface roughness increases, the traces broaden and become approximately centered on the origin, indicating a crossover toward the disorder-dominated regime observed experimentally\cite{volmer2026mapping}. Finally, the fluctuations of the valley splitting around its mean exhibit statistically significant negative autocorrelations [Fig.~\ref{fig:speedvsvptracescov}(e), for $\Delta x\geq\lambda_{\mathrm{max}}=\SI{100}{\nano\meter}$] arising from the long-wavelength properties of the considered rough interfaces. Such anti-correlations are not expected within the effective mass approximation~\cite{david2024long} and drive increased valley leakage over long shuttling distances, thereby highlighting the relevance of a detailed description of atomistic disorder for shuttling simulations.

Finally, having obtained time traces of the valley splitting $E_{\mathrm{v}}$ and valley phase $\phi_{\mathrm{v}}$ [Fig.~\ref{fig:speedvsvptracescov}(b)--(c)], we may now estimate leakage into the valley-excited state as a function of shuttling speed. The shuttled electron state may be projected onto a relevant 2D subspace as
\begin{equation}
\ket{\psi(t)}=\alpha_{0}(t)\ket{\psi_{0}(t)}+\alpha_{1}(t)\ket{\psi_{1}(t)}\,,
\label{eq:schrod_psi}
\end{equation}
where $\psi_{m}(\mathbf{r},t)=\braket{\mathbf{r}|\psi_{m}(t)}$ for $m\in\{0,1\}$. In this subspace, the time-dependent Schr\"odinger equation may be expressed as~\cite{friesen2007valley,nagai2025digital}
\begin{equation}
i\hbar \frac{d}{dt}
\begin{bmatrix}
\alpha_{0}(t) \\
\alpha_{1}(t)
\end{bmatrix}
=
\frac{1}{2}
\begin{bmatrix}
- E_{\mathrm{v}}(t) & \hbar\dot{\phi}_{\mathrm{v}}(t) \\
\hbar\dot{\phi}_{\mathrm{v}}(t) & E_{\mathrm{v}}(t)
\end{bmatrix}
\begin{bmatrix}
\alpha_{0}(t) \\
\alpha_{1}(t)
\end{bmatrix},
\label{eq:schrod}
\end{equation}
where $\hbar$ is the reduced Planck constant and $\dot{\phi}_{\mathrm{v}}(t)$ is the time-derivative of the valley phase. Denoting the interpolations of the valley splitting and valley phase [Fig.~\ref{fig:speedvsvptracescov}(b)--(c)] as $E^{\mathrm{interp}}_{\mathrm{v}}(t/T)$ and $\phi^{\mathrm{interp}}_{\mathrm{v}}(t/T)$, respectively, the time traces of the valley splitting and valley phase derivative may be expressed in terms of the average shuttling speed $v=L/T$,
\begin{equation}
E_{\mathrm{v}}(t) \approx E^{\mathrm{interp}}_{\mathrm{v}}\left(\frac{t}{T}\right) \,, \quad \dot{\phi}_{\mathrm{v}}(t) \approx \frac{v}{L} \dot{\phi}^{\mathrm{interp}}_{\mathrm{v}}\left(\frac{t}{T}\right)\,.
\end{equation}
We note that Eqs.~\ref{eq:schrod_psi} and \ref{eq:schrod} explicitly capture non-adiabatic, coherent valley dynamics arising from atomistic disorder, which is known to be a dominant cause of valley mixing and excitation during shuttling~\cite{losert2024strategies,david2024long,volmer2025reduction}, as well as strain inhomogeneity. In contrast, dissipative valley relaxation processes---notably those due to electron--phonon scattering~\cite{yang2013spin}, pulse imperfections, and charge defects~\cite{jeon2025robustness}---are not included, and are thus ignored in the derived shuttling fidelities.

Initializing the electron in the ground state $\ket{\psi(t=0)}=\ket{\psi_{0}(t=0)}$ we solve Eq.~\ref{eq:schrod} by discretizing time, approximating the Hamiltonian as constant over each time step. We estimate the shuttling fidelity as
\begin{equation}
\mathcal{F}=\min_{t\in[0,T]} \left|\alpha_{0}(t)\right|^2\,.
\end{equation}
A fidelity of $\mathcal{F}=1$ would correspond to a perfect preservation of the valley degree of freedom of the electron during shuttling. Although quantum information is encoded in the spin degree of freedom, gate operations and readout assume the electron remains in the valley-ground state. Loss of valley fidelity therefore undermines reliable spin measurements and multi-qubit operations~\cite{buterakos2021spin}. An explicit treatment of spin, spin--orbit coupling, spin--valley coupling, and orbital leakage could be achieved by constructing and propagating the full spin-resolved, time-dependent atomistic TB Hamiltonian using Trotterized time evolution, albeit at increased computational cost.

\begin{figure}[!tbp]
\includegraphics[width=\columnwidth]{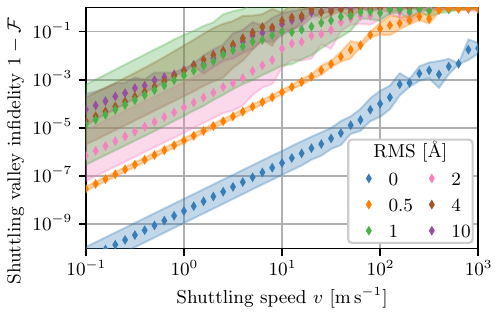}
\caption{Shuttling infidelity due to leakage into the valley-excited state for electrons shuttled in CBs with rough Si--SiGe interface RMS heights ranging from $0$ to $\SI{10}{\angstrom}$. The shaded areas indicate the variability of the infidelity, corresponding to $\pm 1$ standard deviation computed from ensembles of six ($\mathrm{RMS}=\SI{1}{\angstrom}$) or three (other RMS heights) CB atomic structures.}
\label{fig:fidelity}
\end{figure}

In Fig.~\ref{fig:fidelity}, we plot the shuttling infidelity $1-\mathcal{F}$ as a function of shuttling speed. The infidelity is seen to generally grow with the rough Si--SiGe interface RMS height, reflecting both the higher leakage rate ($\dot{\phi}_{\mathrm{v}}$) and smaller gap ($E_{\mathrm{v}}$) in CBs with high disorder; the spread in fidelities likewise increases with disorder. A notable exception occurs for RMS heights close to the interatomic layer spacing, \SI{1}{\angstrom}, where destructive interferences in the valley coupling collapse the valley splitting [Fig.~\ref{fig:speedvsvptracescov}(b)] and thereby drive leakage into the valley-excited state [Eq.~\ref{eq:schrod}].

Overall, we demonstrate that shuttling fidelities degrade markedly with increasing interface roughness in Si--SiGe CB device, with a notable anomaly near the atomic-layer scale. This result emerges from a multiscale simulation framework that notably combines time-dependent finite-element electrostatics with atomistic tight-binding simulations, enabling faithful treatment of both micrometer-scale device geometry and angstrom-scale valley physics. Our approach establishes clear quantitative constraints on tolerable disorder, thereby providing practical design guidelines for quantum engineers aiming to scale shuttling architectures. While we focused on conventional Si--SiGe heterostructures, our multiscale simulation framework may be extended to emerging CB platforms, notably Si--SiO$_2$ devices, where interface roughness modeling is especially needed, as well as engineered quantum wells that maximize valley splitting, such as wiggle wells~\cite{losert2023practical}. Moreover, our framework is general; it naturally extends to other disorder sources, such as Ge concentration gradients~\cite{paquelet2022atomic}, charge defects~\cite{paz2024fdsoi,ciroth2025numerical}, misfit dislocations~\cite{liu2023strain}, strain induced by metallic gates~\cite{frink2025reducing}, and crosshatch patterns~\cite{albrecht1995surface}. Recent work has also investigated spin dephasing in the presence of sparse negatively charged defects~\cite{ciroth2025numerical}; within an extension of our framework, trapping by positively charged defects could in principle be modeled by describing the shuttle-localized and defect-localized electron states (and their tunnel coupling) within the atomistic tight-binding framework, while irreversible capture would additionally require dissipative relaxation processes. These directions open the way toward systematic and atomistically-informed strategies to achieve robust and scalable shuttling architectures.

We gratefully acknowledge financial support from the QSP 015 project within the ``Internet of Things: Quantum Sensors Challenge'' Program (QSP) of the National Research Council (NRC) of Canada.

\section*{Author Declarations}
\subsection*{Conflict of Interest}
F\'{e}lix Beaudoin is the chief executive officer of Nanoacademic Technologies Inc. and owns equity in the company.
\subsection*{Author Contributions}
\textbf{R. J. Prentki:} Conceptualization (equal); methodology (equal); software (lead); investigation (lead); data curation (lead); formal analysis (lead); validation (lead); visualization (lead); writing---original draft preparation (lead); writing---review and editing (lead).
\textbf{P. Philippopoulos:} Conceptualization (equal); methodology (equal); software (supporting); writing---review and editing (equal).
\textbf{M. R. Mostaan:} Conceptualization (supporting); methodology (supporting); software (supporting); writing---review and editing (supporting).
\textbf{F. Beaudoin:} Funding acquisition (lead); conceptualization (equal); methodology (supporting); software (supporting); project administration (lead); writing---review and editing (equal).

\section*{Data Availability Statement}
The data that support the findings of this study are available from the corresponding author upon reasonable request.

%

\end{document}